\documentclass[twocolumn, prl]{revtex4}
\pdfoutput=1
\usepackage{graphicx}
\usepackage{epsfig}
\begin{document}
\title{Reponse to Comment on ``Direct mapping of the finite temperature phase diagram of strongly correlated quantum models" }
\author{Qi Zhou$^{1}$, Yasuyuki Kato$^{2}$, Naoki Kawashima$^{2}$, and Nandini Trivedi$^{3}$}
\affiliation{$^{1}$ Joint Quantum Institute, University of Maryland, College Park, MD 20742\\
$^{2}$Institute for Solid State Physics, University of Tokyo, Kashiwa, Chiba 277-8581, Japan \\
$^{3}$Department of Physics, The Ohio State University, Columbus, OH 43210}

\maketitle

Pollet et al have questioned our claim staked in ~\cite{ourpaper} that from the location of sharp features in the compressibility $\kappa$ in the trap, it is possible to map out the phase boundary 
between two phases within reasonable accuracy. We stand by it and use this reply to sharpen the conditions under which ``sharp" features in $\kappa$ reflect the effects of critical fluctuations.

(1) The first point raised by Pollet et al ~\cite{Bcomment} is that sharp density gradients 
may also lead to sharp features in $\kappa$.
We agree both critical fluctuations and strong density gradients, in general, can contribute to the features in $\kappa$, as seen in Fig. 4 of our PRL. However, it is easy to separate out effects due to density gradients
by simply making the trap potential flatter or by making the system larger by increasing the number of particles. 
The effects due to density gradients will then be diminished. It can also be checked theoretically, by closely corroborating the kink features, with the superfluid density distribution using the local density approximation, as was done in Fig.~5 of our PRL.

(2) The second point raised by Pollet et al is that when they simulate the parameters in our Fig.(5B) in a trap,
they observed no features in $\kappa$ associated with the critical point by showing $\kappa$ within a small region of about 7 lattice spacings near the critical point $r_c$ in their comment. We do not dispute their numerical data but we disagree with their analysis and interpretation.  

\begin{figure}[tbp]
\includegraphics[width=1.6in]{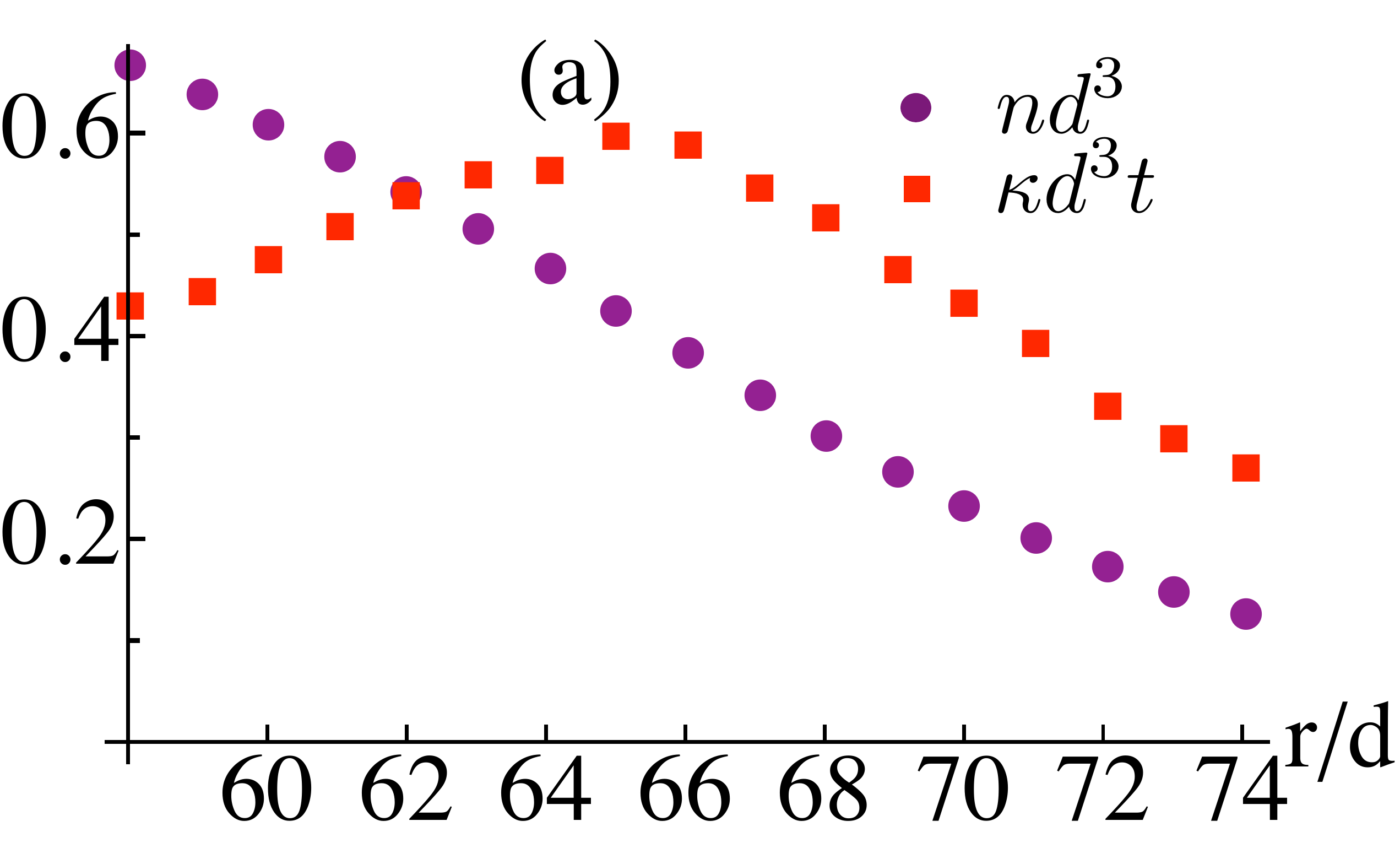}
\includegraphics[width=1.6in]{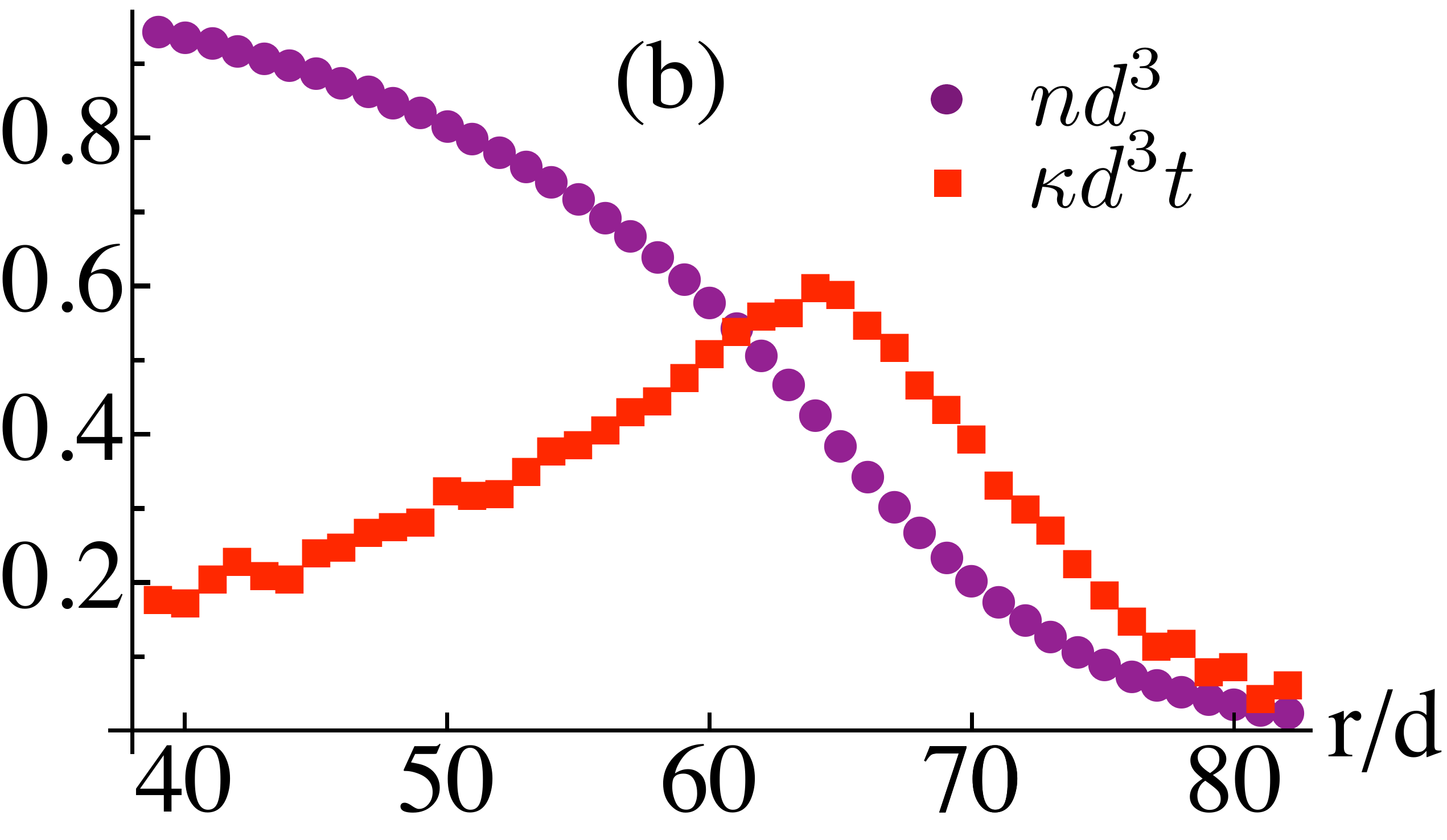}
\includegraphics[width=1.6in]{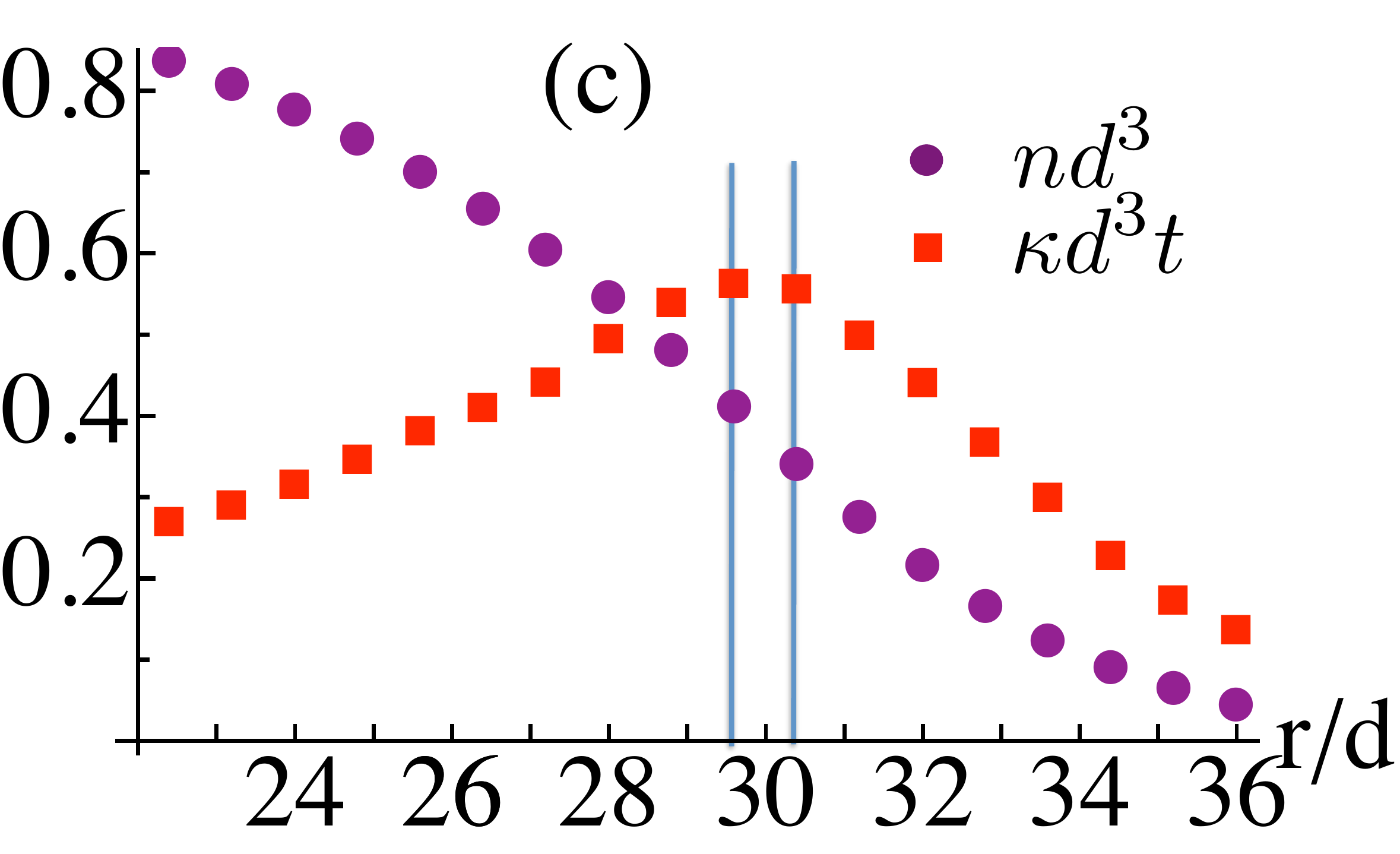}
\includegraphics[width=1.6in]{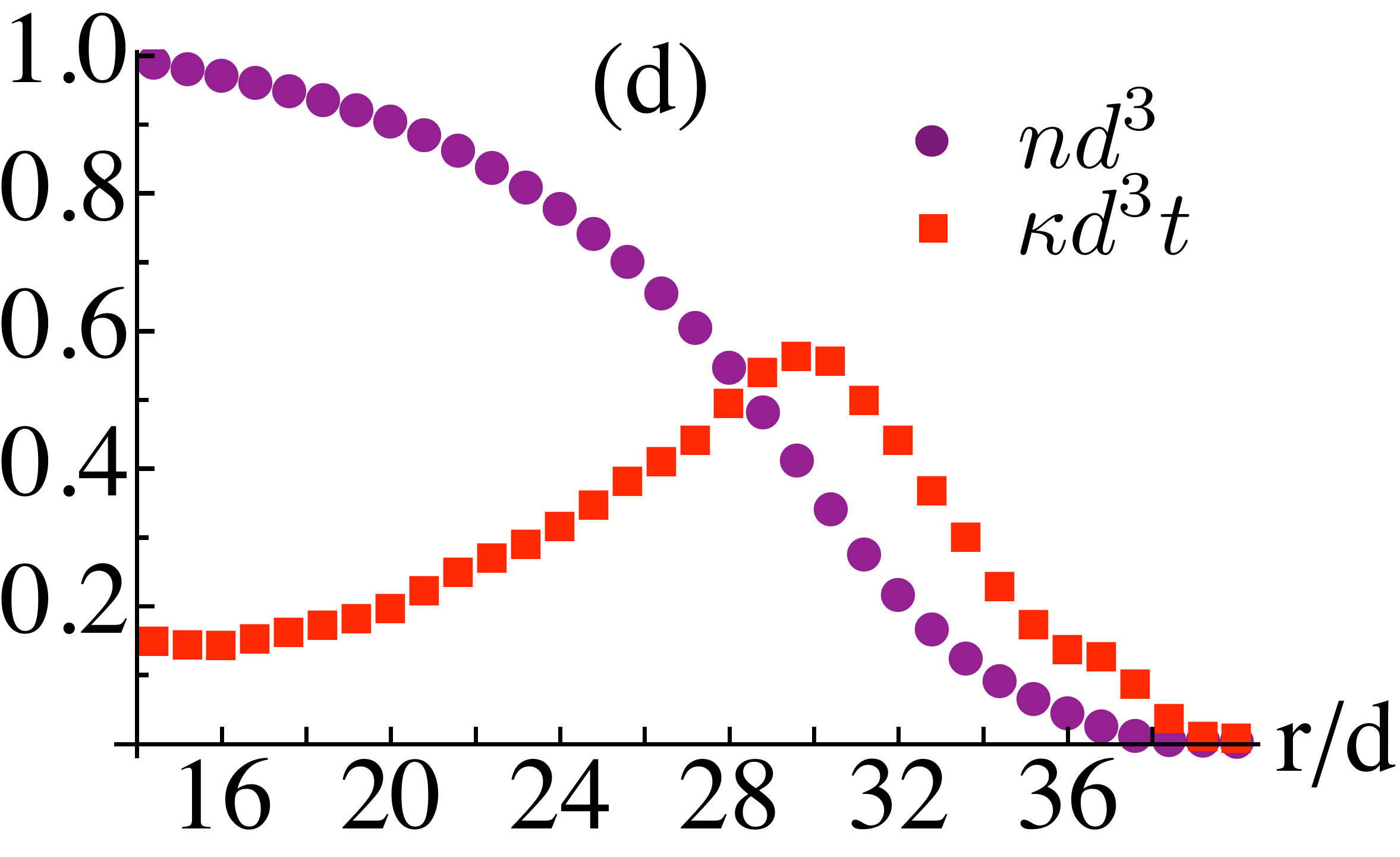}
\caption{(a-b)Re-plot, from Fig.~1 of the comment~\cite{Bcomment}, of the trap density profile $n$ and local compressibility $\kappa$ over a larger scale. (c-d)Our direct simulations in a trap for the same $T$ and $t/U$ as in (a-b) but at different trapping frequency $V/t=0.00262$ and chemical potential $\mu/t=1.984$. }
\end{figure}

We are well aware that singular features in $\kappa$ will get rounded by finite size effects. However, we want to make a definite point here, that even with this rounding, if the data is analyzed carefully, it should be possible to locate the phase boundary within reasonable accuracy. 

We first demonstrate this fact using the {\em same} data generated by Pollet et al. In Fig.~1(a), We re-plot their data in a larger region, up to 17 lattice spacings, than what has been shown in their comment. It is now clear that there is a maximum in $\kappa$ from which $r_c$ can be easily located. On a larger scale shown in Fig.~1(b), the structure of $\kappa$ becomes even clearer. 

Next we address the question about the accuracy of the method for extracting the phase boundary $\mu_c$ from the maximum in $\kappa$. A quantitative determination of $\mu_c$ and the associated error bar can only
be obtained from a detailed finite size scaling analysis, however, a qualitative estimate can be made by the procedure
outlined below. In Fig.~1(c-d), we show our quantum Monte Carlo simulations in the trap. The finite size rounding in a trap gives an uncertainty of $r_c/d$ between $29.6$ and $30.4$, where $d$ is the lattice spacing. This leads to an uncertainty in the critical chemical potential $\mu_c/U=-0.094\pm0.016$. 
Our result for the mean value of $\mu_c$ is within $2\%$ compared with the mean value of  $\mu_c$ obtained in~\cite{Bpaper} at the same $T/t$ and $t/U$.  The critical density $n_c$ located as  $0.375\pm 0.035$
is also within $1.4\%$ of the mean value of $n_c$ found in~\cite{Bpaper}.  

Thus based on this comparison with Pollet et al's own results, we emphasize that the finite temperature critical point of quantum models can be identified within a reasonable accuracy from the derivative of the density profile of atoms in the trap, despite the finite size rounding effect in the vicinity of the critical point. 

We also use this opportunity to correct typos in our PRL pointed out by Pollet et. al:
The $x$ label of Fig.(4A) should be $10, 20, 30,40$ and also the triangle and box symbols should be reversed. The label of $\mu/U$ for Fig.(5B) should be $0.5, 0.448, 0.290, 0.028, -0.339$.


\begin{thebibliography}{99}
\bibitem{ourpaper} Q. Zhou, Y. Kato, N.Kawashima and N.Trivedi, Phys. Rev. Lett, 103, 085701(2009)
\bibitem{Bcomment} Comment by L.Pollet, et.al.
\bibitem{Bpaper} L.Pollet, N.V., Prokof'ev and B.V. Svistunov, arXiv:1003.2655v2(2010)

\end{thebibliography}
\end{document}